\documentclass[%
 reprint,
 superscriptaddress,
 amsmath,amssymb,
 aps,
 prl,
 longbibliography,
lengthcheck,%
]{revtex4-1}

\usepackage{graphicx}
\usepackage{dcolumn}
\usepackage{bm}
\usepackage{color}
\usepackage{ulem}
\usepackage{hyperref}
\usepackage{braket}
\usepackage{physics}
\usepackage{xr}

\begin{document}

\preprint{APS/123-QED}

\title{
Enhancement of the Topological Hall Effect through Engineering the Skyrmion Size and Shape
}

\author{Kazuma Murata}
\affiliation{
Department of Physics, Institute of Science Tokyo, Meguro, Tokyo, 152-8551, Japan
}

\author{Ryunosuke Terasawa}
\affiliation{
Department of Physics, Institute of Science Tokyo, Meguro, Tokyo, 152-8551, Japan
}

\author{Hiroaki Ishizuka}
\affiliation{
Department of Physics, Institute of Science Tokyo, Meguro, Tokyo, 152-8551, Japan
}

\date{\today}

\begin{abstract}
The topological Hall effect of skyrmions is a promising electrical probe of magnetic textures, but its small magnitude limits applications.
In this work, we theoretically investigate the topological Hall effect induced by sparsely distributed skyrmions in the weak-coupling limit, focusing on how the size and shape of the skyrmion affect the Hall effect.
To this end, we used a formalism based on angular-momentum eigenstates, which enables systematic analysis within reasonable computation time. We found that, in clean materials where the electron mean-free path is comparable to or larger than the skyrmion size, the Hall effect depends considerably on the size and shape of the skyrmion.
In particular, the Hall effect is largest when $k_F\lambda\sim1$, and its magnitude can be tuned by the domain-wall width of skyrmions.
These findings provide a basis for investigating how the shape and size of skyrmions affect transport properties and demonstrate that the topological Hall effect is controllable by engineering these properties.
\end{abstract}

\pacs{
}

\maketitle

\begin{figure}
  \includegraphics[width=\linewidth]{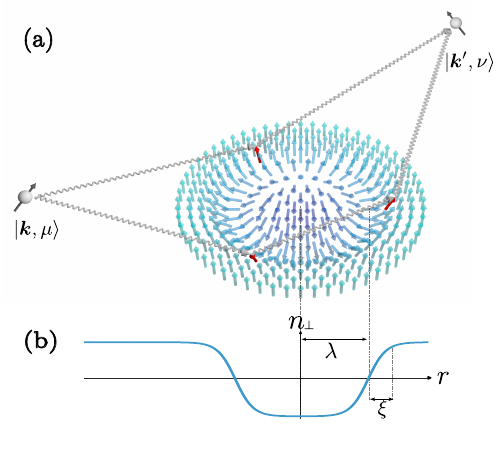}
  \caption{
    (a) A schematic of electron scattering by a magnetic skyrmion.
    An incident electron in the $\lvert \bm{k},\mu\rangle$ state is scattered into the $\lvert \bm{k}',\nu\rangle$ state by the exchange interaction.
    The red arrows indicate spins at which electrons are scattered, and the gray lines are the trajectories of single- and multiple-spin-scattering paths that give rise to the asymmetric scattering. 
    (b) Radial profile of the out-of-plane spin component $n_\perp(r)$ for the skyrmion in Eq.~\eqref{model:Skyrmion_solution}. 
    The skyrmion radius $\lambda$ corresponds to the radial position at which $n_\perp(r)=0$, and $\xi$ characterizes the width of the domain wall.
}
\label{fig:model}
\end{figure}

{\it Introduction} ---
%
%
Magnetic materials host rich magnetic textures, which give rise to unique material properties~\cite{Fert2017a,Tokura2021a}.
Magnetic skyrmion~\cite{Belavin1975a,Bogdanov1989a}, a swirling magnetic defect shown in Fig.~\ref{fig:model}(a), is one such magnetic texture that has been studied 
in a wide variety of materials~\cite{romming2013writing,hagemeister2015stability,yu2010real,muhlbauer2009skyrmion} and theoretically in various magnetic systems \cite{okubo2012multiple,leonov2015multiply,schutte2014dynamics}.
It is a topological defect that is stable once created, and in magnetic metals, skyrmions can be moved by electric current.
Such mobility makes it particularly interesting for applications, such as magnetic memories~\cite{Kang2016a,Blasing2020a}.

%
%
In metals, chiral magnetic textures often affect transport properties, enabling detection of magnetic states and control of material properties through controlling magnetism.
A key concept in such phenomena is the chirality of the magnetic state: 
An anomalous Hall effect occurs without the spin-orbit interaction if a magnetic state has the scalar spin chirality $\bm S_i\cdot\bm S_j\times\bm S_k$~\cite{Ye1999a,Ohgushi2000a,Tatara2002a}, a phenomenon often referred to as the topological Hall effect (THE), and a nonreciprocal current related to the vector spin chirality $\bm S_i\times\bm S_j$~\cite{Ishizuka2020a}.
The THE has been observed in a variety of materials, from transition metal compounds with non-collinear magnetic anisotropy~\cite{Taguchi2001a,Machida2007a} to frustrated magnets~\cite{Takatsu2010a,Ok2013a,Nakatsuji2015a,Mayoh2022a,Takagi2023a,Okuma2024a}, as well as
in skyrmion materials~\cite{Ye1999a,Onoda2004a,Neubauer2009a,Kanazawa2011a}. 
The nonreciprocal current was reported in helical magnets~\cite{Yokouchi2017a,Aoki2019a,mayo_band_2025}.
These phenomena detect the chiral magnetic state and are therefore potentially applicable to reading out magnetic memories using magnetic textures like skyrmions~\cite{Blasing2020a}.
However, the voltage induced by these chirality-related phenomena is often small in materials with large or dilute skyrmions, posing a challenge for their application~\cite {Maccariello2018a}.

%
%
In this work, we study the possibility of tuning THE by controlling the skyrmion shape.
A recent study has pointed out that, in highly conductive materials, the scalar spin chirality of further-neighbor spins has a considerable effect on the THE by skew scattering~\cite{Ishizuka2018a,Terasawa2025a}.
The further-neighbor contribution may give rise to size- and shape-dependent THE, in contrast to the intrinsic contribution.
However, understanding of such effects has been limited due to the computational complexity.
In this work, we studied how the shape and size of a skyrmion affect the THE by deriving a formula for the Hall conductivity based on the angular-momentum eigenstates. 
We find that the THE shows a pronounced peak at $k_F\lambda\sim1$ where $k_F$ and $\lambda$ are the Fermi wavenumber and skyrmion radius, respectively.
The maximum Hall conductivity can be further tuned by increasing the domain-wall width $\xi$, which reaches one order of magnitude enhancement in the thick wall case.
We also demonstrate that the sign and magnitude of the THE can be controlled by tuning $k_F\lambda$ and $\xi/\lambda$.
These results demonstrate that the THE in highly conductive magnets is highly tunable with skyrmion size and shape, enabling control of the THE in magnetic devices.

{\it Model} --- 
We consider a two-dimensional itinerant electron system coupled to 
localized magnetic moments.
The Hamiltonian consists of the kinetic term and the Kondo coupling
\begin{align}
H&=
\sum_{\bm{k},\mu}
\frac{k^2}{2m^\ast}
c_{\bm{k},\mu}^{\dagger}c_{\bm{k},\mu}
-J\sum_{i,\mu,\nu}\bm{S}_{i}\cdot
c_{i,\mu}^{\dagger}
\bm{\sigma}_{\mu\nu}
c_{i,\nu}.
\label{eq:model_H}
\end{align}
Here, $c_{\bm{k},\mu}^\dagger$ is the creation operators for an electron with spin $\mu$ and wave vector $\bm{k}$, $c_{i,\mu}^{\dagger}$ is that for an electron with spin $\mu$ on site $i$, $J$ is the Kondo-coupling constant, $\bm{S}_{i}$ is the local classical spin at site $i$, and $\bm{\sigma}_{\mu\nu}=(\sigma_{\mu\nu}^{x},\sigma_{\mu\nu}^{y},\sigma_{\mu\nu}^{z})$ is the vector composed of the matrix elements of the Pauli matrices. 
The operators $c_{\bm{k},\mu}$ and $c_{i,\mu}$ are related by $c_{i,\mu}=\frac{1}{\sqrt{N}}\sum_{\bm{k}}e^{i\bm{k}\cdot\bm{R}_i}c_{\bm{k},\mu}$, where $N$ is the number of sites.

To study the Hall effect by isolated skyrmions, we consider a skyrmion with skyrmion number $q$ at position $\bm r=\bm 0$.
The spin at $\bm{r}=r(\cos\theta,\sin\theta)$ is given by
\begin{align}
\bm{S}(\bm{r})
=
S
\begin{pmatrix}
n_{\parallel}(r)\cos(q\theta+\phi_{0})\\
n_{\parallel}(r)\sin(q\theta+\phi_{0})\\
n_{\perp}(r)
\end{pmatrix}.
\label{eq:model_skyrmion}
\end{align}
In the following, we mainly consider a skyrmion given by
\begin{align}
    \begin{aligned}
    n_\perp(r)&=\tanh\qty(\frac{r-\lambda}{\xi}),\\
    n_\parallel(r)&=\text{sech}\qty(\frac{r-\lambda}{\xi}),
    \end{aligned}\label{model:Skyrmion_solution}
\end{align}
unless stated otherwise. Here, $\lambda$ is the skyrmion radius and $\xi$ is the domain-wall width
[see Fig.~\ref{fig:model}(b)].

{\it Spin texture and Hall effect} ---
To study the THE, we focus on the skew scattering contribution~\cite{Ishizuka2018a,Ishizuka2021a}, which is expected to be dominant in clean materials~\cite{Nagaosa2010a}.
Skew scattering is a process in which a Hall effect arises from the left--right asymmetry in electron scattering rate~\cite{Smit1958a}, which is formally expressed as the antisymmetric component of the scattering rate, $w^{(-)}=\qty(W_{n\to m}-W_{m\to n})/2$, where $W_{n\to m}$ is the scattering rate for an electron to be scattered from a state $\ket{n}$ to a state $\ket{m}$ by a scattering potential.
Applying the second Born approximation to the magnetic textures in a ferromagnetic background, assuming the coupling constant $J$ is small, the Hall conductivity for the Hamiltonian in Eq.~\eqref{eq:model_H} reads~\cite{Ishizuka2018a,Terasawa2025a},
\begin{align}
  \sigma_{xy} &= 
  \frac{\tau^2e^2S^3a^2J^3|m^\ast|L^2}{\pi N^2}k_{\text{F}}^2
  \sum_{i, j, h}
  {\delta\bm{n}_{h}\cdot(\delta\bm{n}_{i}\times \delta\bm{n}_j)}\notag\\
  &\,\,
  J_{0}(k_{\text{F}}R_{ij})J_{1}(k_{\text{F}}R_{jh})J_{1}(k_{\text{F}}R_{ih})\hat{\bm{z}}\cdot(\hat{\bm{R}}_{ih}\times\hat{\bm{R}}_{jh}).\label{eq:Hall_conductivity}
\end{align}
Here, $\tau$ is the relaxation time, $k_{\text{F}}$ is the Fermi wavenumber, $m^\ast$ is the effective mass of the electron, $a$ is the lattice constant, and $L^2$ is the system size.
We neglect the spin splitting of the electronic bands, as it gives only an $\mathcal{O}(J^4)$ correction to the Hall conductivity in Eq.~\eqref{eq:Hall_conductivity}.
Note that Eq.~\eqref{eq:Hall_conductivity} is applicable to an arbitrary spin configuration $\bm{n}(\bm r)$.
The derivation is in Sec.\uppercase\expandafter{\romannumeral 1} of the Supplemental Material~\cite{suppl}.

In a system with dilute, isolated skyrmions, in which $k_Fa\ll1$ and $\delta\bm n$ changes slowly, Eq.~\eqref{eq:Hall_conductivity} reads
\begin{align}
\sigma_{xy}
&=-2\pi^3\sigma_{0}k_{F}^2\sum_{m=-\infty}^{\infty}F_{m,q},
\label{eq:Hall_conductivity_1}\\
F_{m,q}&=\Big( N^{(\perp)}_{m+1+q,m+1+q}-N^{(\perp)}_{m+q,m+q}-N^{(\perp)}_{m+1,m+1} \notag\\
&\quad\quad+ N^{(\perp)}_{m,m} \Big)
\left(  N^{(\parallel)}_{m+1+q,m+1}- N^{(\parallel)}_{m+q,m} \right)^2,\\
&=F_{-q-1-m,q},\notag\\
N^{(\perp)}_{m,l}&=\int_{0}^{\infty} dr\,\,r\qty(n_{\perp}(r)-1)J_{m}(k_{\text{F}}r)J_{l}(k_{\text{F}}r),\\
N^{(\parallel)}_{m,l}&=\int_{0}^{\infty} dr\,\,rn_{\parallel}(r)J_{m}(k_{\text{F}}r)J_{l}(k_{\text{F}}r),
\end{align}
where $\sigma_0 =\frac{\tau^2e^2S^3J^3|m^\ast|}{\pi}n_\text{sk}$; $m^\ast>0$ corresponds to the $n$-type carriers and $m^\ast<0$ is the $p$-type carriers (see Sec.\uppercase\expandafter{\romannumeral 2} of the Supplemental Material~\cite{suppl}).
Here, $n_\text{sk}$ is the density of skyrmions in the system and $n_{\parallel}(r)$ and $n_{\perp}(r)$ are functions satisfying $n_{\perp}(r)^{2}+n_{\parallel}(r)^{2}=1$, $\lim_{r\to \infty}n_{\perp}(r)=1$, and $ \lim_{r\to \infty}n_{\parallel}(r)=0$.

The integrals $N^{(\perp)}_{l,l}$ and $N^{(\parallel)}_{l+q,l}$ 
correspond to
\begin{align}
N^{(\perp)}_{l,l}&=\bra{k_{\text{F}},l,\mu}H_K\ket{k_{\text{F}},l,\mu}\\  N^{(\parallel)}_{l+q,l}&=\bra{k_{\text{F}},l+q,\downarrow}H_K\ket{k_{\text{F}},l,\uparrow}, 
\end{align}
where $\ket{k_{\text{F}},l,\mu}$ is the electron eigenstate in cylindrical coordinates with wavenumber $k_{\text{F}}$, angular momentum $l$, and spin $\mu$.
These equations show that $N^{(\perp)}_{m,l}$ and $N^{(\parallel)}_{m,l}$ are the scattering strengths of the spin-conserving and spin-flip processes, respectively;
only $N^{(\perp)}_{l,l}$ and $N^{(\parallel)}_{l+q,l}$, and not the other subscripts, appear in Eq.~\eqref{eq:Hall_conductivity_1} as $q$ corresponds to the change of angular momentum by the scattering.
$m=l$ and $l=m\pm q$, respectively.
Mathematically, it is related to a commutation relation $[\mathcal{J}_z,H]=0$ in the continuum limit where $\mathcal{J}_z=-i\partial_\theta+\frac{q}{2}\sigma^z$, i.e., the conservation of $\mathcal{J}_z$. 
The Hall conductivity in Eq.~\eqref{eq:Hall_conductivity_1} implies that the skyrmion number $q$ is incorporated into the scattering process through the orbital angular momentum transferred upon electron spin flip. 
For the skyrmion solution in Eq.~\eqref{model:Skyrmion_solution}, the Hall conductivity reads $\sigma_{xy}=\lambda^{4}\tilde{\sigma}_{xy}$, where
\begin{align}
\tilde{\sigma}_{xy}&=-2\pi^3\sigma_0(k_F\lambda)^2
\sum_{m=-\infty}^{\infty}\tilde{F}_{m,q},\notag\\
\tilde{F}_{m,q}&=\Big( \tilde{N}^{(\perp)}_{m+1+q,m+1+q}-\tilde{N}^{(\perp)}_{m+q,m+q}-\tilde{N}^{(\perp)}_{m+1,m+1} \notag\\
&\quad\quad+ \tilde{N}^{(\perp)}_{m,m} \Big)
\left(  \tilde{N}^{(\parallel)}_{m+1+q,m+1}- \tilde{N}^{(\parallel)}_{m+q,m} \right)^2
\end{align}
and
\begin{align}
\tilde{N}^{(\perp)}_{m,m}\qty(k_F\lambda,\frac{\xi}{\lambda})&=\int_{0}^{\infty} dx\,\,x\qty{\tanh\qty(\frac{\lambda}{\xi}(x-1))-1}\notag\\
&\qquad  \qquad \times J_{m}(k_{\text{F}}\lambda x)^2,\label{eq:Nperp}\\
\tilde{N}^{(\parallel)}_{m+q,m}\qty(k_F\lambda,\frac{\xi}{\lambda})&=\int_{0}^{\infty} dx\,\,x\text{sech}\qty(\frac{\lambda}{\xi}\qty(x-1)) \notag\\
&\qquad\qquad\times J_{m+q}(k_{\text{F}}\lambda x)J_{m}(k_{\text{F}}\lambda x).\label{eq:Nparallel}
\end{align}
Here, we introduced a dimensionless coordinate $r=\lambda x$.
The rescaled conductivity $\tilde\sigma_{xy}$ depends only on the two dimensionless parameters $k_F\lambda$ and $\xi/\lambda$.

\begin{figure*}[tbp]
  \includegraphics[width=\linewidth]{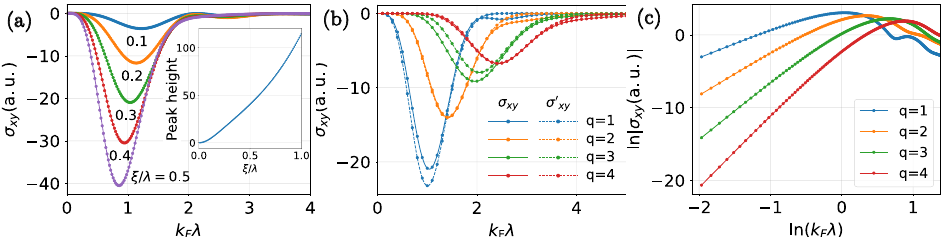}
  \caption{
    Fermi wavenumber dependence of the Hall conductivity $\sigma_{xy}$.
    (a) The $k_F$ dependence of $\sigma_{xy}$ for $q=1$ and $\lambda=1$ with different  $\xi/\lambda$.
    The inset shows the $\xi/\lambda$ dependence of the maximum of $\sigma_{xy}$ in (a).
    (b) The $k_F$ dependence of $\sigma_{xy}$ for $\lambda=1,\xi/\lambda=0.3$ with different $q$. The dashed lines are an approximation in Eq.~\eqref{eq:Hall_conductivity_center}, which gives an accurate description of the peaks of $\sigma_{xy}$. 
    (c) Same data in the logarithmic plot. 
    }
  \label{kFdep}
\end{figure*}

{\it Hall effect by sparsely distributed skyrmions} ---
%
Figure~\ref{kFdep} shows the $k_F$ dependence of $\sigma_{xy}$, obtained by numerically evaluating Eqs.~\eqref{eq:Nperp} and \eqref{eq:Nparallel}.
It shows a non-monotonic $k_F$ dependence with a peak around $k_F\lambda\sim 1$, whose position shifts toward larger $k_F\lambda$ as $q$ increases [Fig.~\ref{kFdep}(b)].
The peak position depends weakly on $\xi/\lambda$, whereas its height shows a strong dependence on $\xi/\lambda$, as shown in Fig.~\ref{kFdep}(a).
It implies that, in a clean material with a long mean free path, the Hall effect can be optimized by tuning $\lambda$ to $k_F\lambda\sim1$, and choosing a material with weak uniaxial anisotropy, which stabilizes a skyrmion with large $\xi$.

In the $k_F\lambda\ll 1$ region, the Hall conductivity scales $\sigma_{xy}\propto k_F^{2q+2}\lambda^{2q+6}$ as in Fig.~\ref{kFdep}(c); it is consistent with the $\sigma_{xy}\propto k_F^{4}$ in the $q=1$ case reported elsewhere~\cite{Terasawa2024a,Mochida2024a}. In contrast, in the $k_F\lambda\gg 1$ region, $\sigma_{xy}$ decays as $k_F^{-4}$ regardless of $q$.
The results show that, unlike the strong coupling limit where the conductivity is related to the skyrmion density~\cite{Ye1999a}, a nontrivial electron-density dependence appears in the weak coupling case.

For the Hall conductivity around the peak at $k_F\lambda\sim1$, we find the results are well approximated by a few terms in Eq.~\eqref{eq:Hall_conductivity_1} [see Fig.~\ref{kFdep}(b)]: $m=-q/2$ and $-q/2-1$ for even $q$, and $m=-(q+1)/2$ for odd $q$.
For instance, for odd $q$, we denote the resulting approximation by $\sigma_{xy}^\prime$, which reads~\cite{suppl},
\begin{align}
&\sigma_{xy}^\prime=
-16\pi^3\sigma_0\lambda^4(k_{F}\lambda)\frac{\lambda}{\xi}I_{2,q}\qty(k_F\lambda,\frac{\lambda}{\xi})I_{1,q}\qty(k_F\lambda,\frac{\lambda}{\xi})^2,
\label{eq:Hall_conductivity_center}\\
&I_{n,q}\qty(y,z)=\int dx\,\sech^{n}\left( z(x-1)\right)
xJ_{\frac{q-1}{2}}(y x)J_{\frac{q+1}{2}}(y x).\notag
\end{align}
Note that $\sech[(\lambda/\xi)(x-1)]$
is a function with a peak around $x=1$ with a width of $\sim\xi/\lambda$, therefore, the integral is dominated by the integrand around $x=1$. This implies that a peak in the $k_F$ dependence appears when $k_F\lambda\sim q/2$, reflecting the maximum of the Bessel function $J_{\nu}(z)$ around $z\sim \nu$. Indeed, S1(b) in the Supplemental Material~\cite{suppl}
shows the peak position is approximately at $0.5q+\text{const}$.

\begin{figure}
  \includegraphics[width=\linewidth]{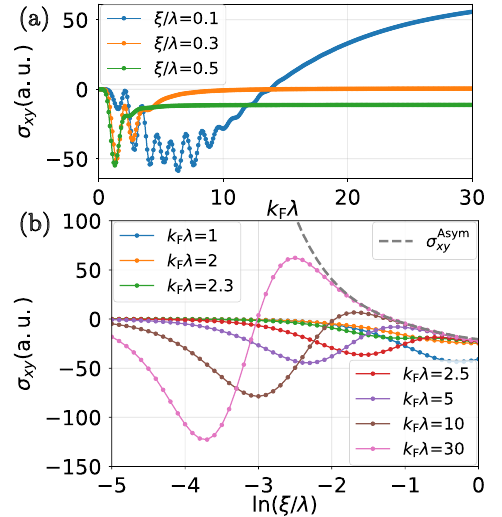}
  \caption{
  (a) $\lambda$ dependence of the Hall conductivity $\sigma_{xy}$ in Eq.~\eqref{eq:Hall_conductivity_1} for $q=1$ and $k_F=1$.
  (b) $\xi$ dependence of $\sigma_{xy}$ for $q=1,k_F=1$ and $k_F\lambda=1,2,2.3,2.5,5,10,30$
  and its asymptotic form $\sigma_{xy}^{(\text{Asym})}$ in Eq.~\eqref{eq:kFlambda_infty_xi/lambda_fixed}. 
  }
  \label{fig3}
\end{figure}

{\it Large skyrmion limit} --- 
Figure~\ref{fig3}(a) shows the $\lambda$ dependence of $\sigma_{xy}$.
In the 
$k_F\lambda\lesssim \lambda/\xi$ regime, which is equivalent to $k_F\xi\lesssim1$, 
the result shows an oscillation of $\sigma_{xy}$.
The oscillation in Fig.~\ref{fig3}(a), absent in Fig.~\ref{kFdep}(a), arises from the $\lambda^4$ prefactor in $\sigma_{xy}=\lambda^4\tilde{\sigma}_{xy}$; similar oscillation was also reported in Ref.~\cite{denisov2017nontrivial}.
Another consequence of the $\lambda^4$ prefactor is the convergence of $\sigma_{xy}$ to a non-zero $\sigma_{xy}$ at $\lambda\to\infty$.
In the $\lambda\to\infty$ limit, the asymptotic form of $\tilde \sigma_{xy}$ reads~\cite{suppl}, 
\begin{align}
&\tilde\sigma^{(\text{Asym})}_{xy}=-4\sigma_{0} \frac{q}{(k_{F}\lambda)^4}\int_{0}^{\infty}d\alpha\,
\frac{\partial^2P(\alpha)}{\partial \alpha^2}\left(\frac{\partial Q(\alpha)}{\partial \alpha} \right)^2\label{eq:kFlambda_infty_xi/lambda_fixed}\\
&P(\alpha)\equiv-\frac{\lambda}{\xi}\int_{\alpha}^{\infty}dx\,\sech^2\left( \frac{\lambda}{\xi}(x-1) \right)\sqrt{x^2-\alpha^2 }\\
&Q(\alpha)\equiv\int_{\alpha}^{\infty}dx\,\sech\left( \frac{\lambda}{\xi}(x-1) \right) \frac{\text{T}_{q}\left( \frac{\alpha}{x} \right)}{\sqrt{ 1-\left( \frac{\alpha}{x} \right)^2 }}
\end{align}
where $\alpha=(m+\frac{q+1}{2})/(k_{F}\lambda)$ and $T_\nu(z)$ is Chebyshev polynomial of the first kind.
The asymptotic form indicates that $\sigma_{xy}\sim\lambda^4\tilde\sigma^{(\text{Asym})}_{xy}\propto \lambda^0k_F^{-4}$, whose coefficient depends only on $\xi/\lambda$ and $q$. 
Namely, the asymptotic form reproduces the converged value 
of the $\lambda$ dependence.
As shown by the dashed line in Fig.~\ref{fig3}(b), 
the convergence of $\lambda$ dependence $\sigma_{xy}^{(\mathrm{Asym})}$ increases as $\xi/\lambda$ decreases, changing sign from negative to positive. 
This is a behavior distinct from the strong-coupling~\cite{Ye1999a} and small $k_F$~\cite{Ishizuka2018a} limits, in which the Hall conductivity does not depend on the shape of the skyrmion.

We next look into the $k_F\xi\lesssim 1$ region.
As shown in Fig.~\ref{fig3}(b), 
the Hall conductivity exhibits a non-monotonic dependence on $\xi/\lambda$ at sufficiently large $k_F\lambda$.
For $q=1$, this non-monotonic behavior emerges for $k_F\lambda\gtrsim2.3$ and is accompanied by two sign reversals for $k_F\lambda\gtrsim10$.
In contrast, for $k_F\lambda\lesssim2.3$, the magnitude of $\sigma_{xy}$ increases monotonically with increasing $\xi/\lambda$.
Similar behavior is also found for the skyrmions with $q>1$ (see Fig.S1(a) of the Supplemental Material~\cite{suppl}).
The results imply that the sign reversal of $\sigma_{xy}$ occurs in a clean material with a skyrmion radius of 10 lattice spacings, provided that $1/k_F$ is of the order of the lattice spacing.
In the large $\xi/\lambda$ regime where $k_F\xi\gg1$, the asymptotic form $\sigma_{xy}^{(\mathrm{Asym})}$ well reproduces the $\xi/\lambda$ dependence of the Hall conductivity, including the sign reversal on the large-$\xi/\lambda$ side. 
Therefore, the position of this sign reversal is independent of $k_F\lambda$ and determined only by $q$.
With increasing $k_F\lambda$, the non-monotonic structure is pushed toward 
smaller $\xi/\lambda$, and in the limit $k_F\lambda\to\infty$, the asymptotic 
form $\sigma_{xy}^{(\text{Asym})}$ reproduces the $\xi/\lambda$ dependence over the entire range.
Consequently, only the single sign reversal depending only on $q$ remains in this limit, with $\sigma_{xy}$ decreasing monotonically with increasing $\xi/\lambda$.
In addition, in the thin-wall limit $k_F\lambda\cdot \xi/\lambda=k_F\xi\ll1$, for an arbitrary fixed value of $k_F\lambda$, the asymptotic analysis presented in Supplemental Material Sec.\uppercase \expandafter{\romannumeral 6} yields
$\sigma_{xy}\propto(\xi/\lambda)^2$.

{\it Small skyrmion limit} --- Finally, we consider the Hall effect for a small skyrmion, whose radius is $k_F\lambda\lesssim1$.
This region corresponds to a semiconductor with small skyrmions or a lightly-doped semiconductor with a small Fermi surface. For a fixed $\xi/\lambda$, the asymptotic form of $\tilde{\sigma}_{xy}$ in the limit $k_F\lambda\to 0$ reads
\begin{align}
\tilde{\sigma}_{xy}&\sim 
32\pi^3\sigma_0q(q+1)^2(q+2)\left( \frac{k_{F}\lambda}{2} \right)^{2q+2}\notag\\
&\quad\times\left( \frac{\xi}{\lambda} \right)^{2q+6}
\text{Li}_{2}\left( -e^{2\frac{\lambda}{\xi}} \right)
\mathrm{Im}\left[ \text{Li}_{q+2}\left( ie^{\frac{\lambda}{\xi}} \right) \right]^2.\label{eq:kFlambda_small_xi/lambda_fixed}
\end{align}

Details of the derivation are in
Sec.\uppercase \expandafter{\romannumeral 5} of the Supplemental Material~\cite{suppl}.
It shows that the scaling with respect to $k_F$ and $\lambda$ is sensitive to the skyrmion number $q$;
the Hall conductivity $\sigma_{xy}$ scales as $k_F^{2q+2}$ with respect to $k_F$ [Fig.~\ref{kFdep}(c)] and as $\lambda^{2q+6}$ with respect to $\lambda$, 
in contrast to the
$q$-independent large-$k_F\lambda$ behavior
$\sigma_{xy}\propto k_F^{-4}\lambda^{0}$. This $q$-dependent power law on $k_F$ is distinct from the strong coupling limit, in which the Hall conductivity
scales as $\sigma_{xy}\propto qk_F^2$.

{\it Summary} --- 
In this paper, we theoretically investigated the skew-scattering topological Hall effect induced by sparsely distributed skyrmions within the scattering-theory framework using the angular-momentum basis. 
Using the formula, we studied the Hall effect by skew scattering in the $\tanh(x)$-type skyrmions. 
We found that, in clean materials where the electron mean-free path is comparable to or larger than the skyrmion size, the topological Hall effect shows rich behavior depending on the size and shape of skyrmions.
In particular, 
$\sigma_{xy}$ is enhanced when the Fermi wavenumber ($k_F$) is $k_F\lambda\sim1$; the height of the peak varies substantially 
with changing the domain-wall width $\xi$. 
By contrast, for a large skyrmion whose radius is $k_F\lambda\gg1$, 
the Hall effect becomes stronger as $\xi/\lambda$ decreases as long as $k_F\xi \gg1$ is satisfied.
These results demonstrate the sensitivity of the Hall effect to the shape and size of skyrmions, suggesting a new way to control the Hall response.

Regarding experiments, 
the pronounced peak found in the Fermi wave number dependence may enhance the visibility of the skew-scattering signal, thereby facilitating its experimental identification and distinction from other Hall-effect contributions. 
A possible candidate for testing this theory is
$\mathrm{Gd_2PdSi_3}$.
This material has been suggested to possess a weak coupling $J$ between the itinerant electron spins and the skyrmion texture and hosts small skyrmions with $\lambda\simeq3.5a$ \cite{kurumaji2019skyrmion}. 
On the other hand, conventional metallic skyrmion systems are expected to lie in the large-$k_F\lambda$ regime, where the Hall conductivity should be well described by the asymptotic expression in Eq.~\eqref{eq:kFlambda_infty_xi/lambda_fixed}. In this regime, the explicit dependence on the skyrmion radius disappears, and the Hall response is determined by $q$, $\xi/\lambda$, and $k_F$.

\acknowledgments
This work is supported by JSPS KAKENHI (Grant No. JP25H00841 and No. JP23K03275) and JST PRESTO (Grant No. JPMJPR2452).

%
%
%
%
\bibliography{ref} 

\end{document}